\documentclass[%
,preprint%
,secnumarabic
,amssymb]{revtex4}
\usepackage{amsmath}%
\usepackage{bm}%
\usepackage{epsf}%

\newif\ifpdf
\ifx\pdfoutput\undefined
\pdffalse 
\else
\pdfoutput=1 
\pdftrue
\fi

\ifpdf
\usepackage[pdftex]{graphicx}
\else
\usepackage{graphicx}
\fi

\ifpdf
\DeclareGraphicsExtensions{.pdf, .jpg}
\else
\DeclareGraphicsExtensions{.eps, .jpg}
\fi

\begin{document}
\title{Measurement of the differential surface and volume excitation probability of medium energy electrons in solids.}

\author{ Wolfgang S.M. Werner\footnote{werner@iap.tuwien.ac.at, \\
fax:+43-1-58801-13499,~tel:+43-1-58801-13462}}%
\affiliation{Institut f{\" u}r Allgemeine Physik,~Vienna University of Technology,
Wiedner Hauptstra\ss e 8--10,~A~1040~Vienna,~Austria
}
\date{\today}%
\begin{abstract} 
A procedure is developed to rigorously decompose experimental loss  spectra of medium-energy electrons reflected from 
solid surfaces into contributions due to surface and volume electronic excitations. 
This can be achieved by analysis of two spectra acquired under different experimental conditions, e.g. measured  at two different
energies and/or geometrical configurations. The input parameters of this procedure comprise the elastic scattering cross
section and the inelastic mean free path for volume scattering.  The (normalized) differential inelastic mean free path
as well as the differential surface excitation probability  are retrieved by this procedure.
Reflection electron energy loss spectroscopy (REELS) data for Si, Cu and Au  are subjected to this procedure  and the retrieved
differential surface and volume excitation probabilities are compared with data from the literature. 
 The results verify the commonly accepted model for medium energy electron transport in solids with unprecedented detail.

\end{abstract}

\maketitle 

\section{Introduction}
The susceptibility of a solid to polarize under the influence of an external electromagnetic perturbation governs many 
physical
phenomena taking place at the surface of a solid and therefore determines important technological properties of
solid materials. The solid state polarizability is characterized by the frequency and momentum dependent dielectric
function $\varepsilon(\omega,q)$. This quantity can be measured by probing a solid surface with elementary particles,
e.g. by photon- \cite{palik,palik1,henke,chantler,chantler1} or electron- \cite{schattschneider} scattering experiments.
A vast number of such experiments has been conducted in the past from which an extensive database of optical data
has been established. Furthermore, with the advent of density functional theory beyond the ground state
\cite{kohndft}, {\em  ab initio} theoretical calculations of optical data have recently become available \cite{wien2k}. 

Nonetheless there still seems to be a need for experimental work in this field since the available datasets are not
always consistent (in particular in the range between the visible and  vacuum-ultra-violet (VUV)-part of the optical
spectrum)  which complicates comparison with theoretical results.  One commonly accepted reason for the inconsistency of
the available datasets is that the early experiments were not always conducted under ultra-high-vacuum (UHV) conditions.
However, there also exist inherent problems with the conventional methods to measure optical data. In photon scattering
experiments, it is difficult to reach the UV-part of the optical spectrum and, moreover, it is not straightforward to
use such experiments to study structures on the sub nanometer scale since photons are hard to focus. The latter
deficiency can be overcome by employing transmission electron energy loss measurements, which are moreover dominated by
the dielectric response in the UV-regime, but even nowadays these experiments are still not routinely conducted under
UHV-conditions and,  moreover, specimen preparation puts quite a  strict limit on the types of nanostructures that can be
investigated with this technique.

Reflection electron energy loss spectroscopy (REELS) measurements have the potential to  bridge this gap. 
Like any charged  particle scattering experiment, such loss spectra are dominated by the UV-response, the
experimental procedure is extremely simple, it is routinely carried out under UHV conditions and can nowadays be
performed with sub-nm  lateral resolution on a large number of instruments on a specimen prepared in an arbitrary way. 

Owing to the energy dependence  of the quasi-elastic  backscattering coefficient that decreases rapidly above several keV
for any material \cite{werqsasia}, such experiments need to be carried out in the medium energy range (several hundred to
several thousands of an eV). Unfortunately this severely complicates quantitative interpretation of REELS spectra 
which are obscured in this energy range by the occurence of multiple surface and volume  excitations.
Furthermore, the particles are multiple elastically scattered  and the elastic scattering cross section exhibits a
pronounced energy and angular dependence.  Therefore, the interaction of the probing particle with the solid is governed
by multiple scattering processes of different types and is consequently very complex. This is the reason why earlier
proposed procedures to extract optical data from REELS measurements have not led to satisfactory results
\cite{toupr35,vicanek,werpia}. Since, on the other hand, such experiments have a great potential for the measurement of
optical data, as outlined above, an attempt to resolve the problems addressed above seems worthwhile. This requires, in a
first step, the development of  a procedure to extract the distribution of energy losses in an individual bulk and
surface excitation from an experimental spectrum, which is dominated by multiple scattering processes of different types.
In a further step these loss distributions need to be converted to optical data.

In the present work, the first step of this ambitious programme is addressed. The elementary interaction characteristics 
are reviewed and it is shown that in  Fourier space, the loss spectrum can be expressed as a bivariate power
series in the bulk and surface scattering distributions in an individual collision. An algorithm to
reverse this bivariate power series using a pair of loss spectra taken under different experimental conditions is
developed. The procedure is succesfully applied to experimental loss spectra of 20 elemental solids. The resulting distribution of
energy losses is presented for Si, Cu and Au and compared with theoretical results based on optical data from various
sources in the literature.

\section{Theorical}
\subsection{Elementary interaction characteristics}
The degrees of freedom of a medium energy electron traversing a solid are subject to fluctuations brought about by the
strong interaction of the probing electron with the ionic and electronic subsystem of  the solid. Owing to the large mass
difference between the incoming electron and the ionic subsystem on one hand and the similarity of the mass of the
probing electron and the electronic subsystem on the other hand, large momentum transfers are accompanied by small energy
losses and vice versa \cite{werqsasia}. Therefore the distinction between {\em elastic}Êand {\em inelastic} scattering is
a meaningful one for medium energy electrons.

For non-crystalline solids, where diffraction effects (coherent scattering) are insignificant \cite{smewerjes137}, the
interaction with the ionic subsystem is adequately described \cite{werqsasia} in terms of an atomic cross section for
elastic scattering $d\sigma_e(\Omega)/d\Omega$, that can be established {\em ab initio} on the basis of a screened
Coulomb
potential for free atoms\cite{jabsalpow}. The solid is modelled  as a random array of scattering centres, and the mean
distance between successive elastic deflection processes, the so--called elastic mean free path $\lambda_e$ is 
the reciprocal of  the differential elastic cross section integrated over the unit sphere multiplied with the density
$N_a$ of scattering centres \cite{werqsasia}.

The inelastic interaction is commonly conceived as a decelaration of the incoming electron by a polarization field set up
by it inside the solid. The susceptibility of the solid to the external perturbation is given in terms of its frequency
($\omega$) and momentum ($q$) dependent dielectric function  $\varepsilon(\omega,q)$. Extensive data for
the dielectric function of solids are available in the literature for zero momentum transfer \cite{palik,palik1,henke,chantler,chantler1}.
The dielectric function for arbitrary momentum transfers  can be obtained by fitting a Drude--Lindhard  type expansion
to such optical data (see e.g., Ref.~\cite{tungpr49}), invoking an appropriate dispersion relation  to extrapolate the
optical loss function  onto the $(\omega,q)$-plane.

In electron  reflection measurements it is obviously not possible to discriminate individual momentum transfers in an
inelastic
process. Rather,  reflection energy loss spectra represent an average over all possible momentum transfers, in
contrast to transmission experiments, which are conducted in the single--scattering regime and allow to discriminate the
momentum transfer by means of angle resolved experiments \cite{schattschneider}.  Therefore, for REELS,  the basic
physical quantity
 governing inelastic scattering deep inside the solid is the differential inverse inelastic mean free path (DIIMFP)
$W_b(T)$, i.e. the distribution of energy losses $T=\hbar\omega$ per unit pathlength in an individual collision. It is
related to the dielectric function $\varepsilon( \omega,q)$ of the solid via the well known formula \cite{landauem}:
\begin{equation}
\label{ediimfp}
W_b(\omega)=\frac{1}{\pi E}\int\limits_{q_-}^{q_+} {\cal I}m\Big\{ \frac{-1}{\varepsilon(
\omega,q)}\Big\}\frac{dq}{q}
\end{equation}
where the subscript "b" indicates bulk inelastic scattering deep inside the solid,
 $E$ is the energy of the incoming electron and $q$ is the momentum transfer which, for parabolic bands, is
confined by $q_-$ and $q_+$ given by:
\begin{equation}
\label{qminmax}
q_\pm=\sqrt{2E}\pm\sqrt{2(E-\omega)}
\end{equation}
In the above expressions as well as in the remainder of this section, atomic units are used, unless indicated otherwise.
The average path length between successive inelastic collisions in the bulk of the solid, the so--called  inelastic
mean free path $\lambda_i$ (IMFP) is a quantity of paramount importance for electron spectroscopy and is the
reciprocal of the DIIMFP integrated over all possible energy losses. The  differential inverse mean free path
normalized to unity area, that will be denoted by $w_b(T)$ in the following,  is related to the unnormalized DIIMFP via
$w_b(T)=\lambda_i W_b(T)$. 

In the vicinity of the surface, both in vacuum and inside the solid, additional modes of the inelastic interaction,
so--called surface excitations, occur, as required by the boundary conditions of Maxwell's equations. Several models to
describe surface excitations have been put forward in the past
\cite{ritchiepr106,sternpr120,echeniquependry,echenique,yubpr,aripr49,tungpr49,dentpra57,vicanek,aminovpr63}. 
The resonance frequency $\omega_s$ of these surface
 modes (indicated by the subscript "s" in the following) is slightly lower than for volume excitations. The
distribution of surface energy losses decays faster with the energy loss than for volume losses: while the distribution of volume
losses tails off slowly towards the maximum energy loss given by the energy $E$ of the probing electron, the typical
energy loss for which the distribution of surface losses drops below a relevant value is much smaller than for bulk
losses, usually it is less than 
50~eV.  Furthermore, surface excitations exhibit a pronounced depth dependence, decaying rapidly with the depth from
the surface both inside the solid and in vacuum. The decay length is of the order of $v/\omega_s\sim 5~{\rm \AA} $,
where $v$ is the speed of the electron. Since this decay length is of the order of, or smaller than, the elastic mean
free path $\lambda_e$, the part of an electrons trajectory passing through  the surface scattering zone is
approximately
rectilinear,
at least  inasmuch as in vacuum the dynamic interaction of the electron with its image charge can be neglected
\cite{aripr49}. Therefore it makes sense to use the differential surface excitation probability $W_s(E,\theta)$ (DSEP) as
basic physical quantity to describe surface excitations, i.e. the integral of the differential mean free path over the
surface scattering zone, where $\theta$ is the polar angle of surface crossing. Tung and coworkers \cite{tungpr49} give
the following expression for the DSEP:
\begin{equation}
\label{etung}
W_s(\omega,\theta,E)=P_s^+(\omega,\theta,E)+P_s^-(\omega,\theta,E)
\end{equation}
where the quantity $P_s^\pm(\omega,\theta,E)$ is defined as
\begin{equation}
\label{edsep}
P_s^\pm(\omega,\theta,E)=\frac{1}{\pi E\cos\theta}\int\limits_{q_-}^{q_+}\frac{|q_s^\pm |dq}{q^3}
{\cal I}m\Big[\frac{(\varepsilon(\omega,q)-1)^2}{\varepsilon(\omega,q)(\varepsilon(\omega,q)+1)}\Big]
\end{equation}
and 
\begin{equation}
\label{etung3}
q_s^\pm =\Big[q^2-\big(\frac{\omega+q^2/2}{\sqrt{2E}}\big)^2\Big]^{1/2}\cos\theta\pm
(\frac{\omega+q^2/2}{\sqrt{2E}})\sin\theta
\end{equation}
The total surface excitation probability $\langle n_s(\theta,E)\rangle$ is obtained from expression (\ref{etung}) by
integrating over the energy loss. Note that this (dimensionless) quantity, that will be refered to as (total) surface
excitation probability (SEP) below, is equal to the average number of surface excitations in a single surface crossing.
The normalized differential surface excitation probability is therefore given by  $w_s(T,\theta)=W_s(T,\theta)/\langle
n_s(\theta,E)\rangle$. 

 To avoid confusion, it is noted that although the same symbol ("$W$" in $W_b$ and $W_s$) is used to denote the
distribution of energy losses  in a  surface and bulk excitation, the physical meaning of these quantities  is quite
different. While the DIIMFP is the volume scattering probability per unit path length and energy, the DSEP represents the
surface excitation probability per unit energy, since this quantity is obtained by  integrating  over the path the
electron takes through the surface  scattering zone\cite{tungpr49}. The normalized distribution of energy losses $w_b(T)$
and $w_s(T)$ are physically equivalent quantities and both have the dimension of reciprocal energy, which is the reason why
the same symbol is chosen for these quantities (see the next section).
 
Examples for the DIIMFP and DSEP for Si and Au are given in Figure~\ref{fdiimfp}. The DIIMFP is shown for two energies
(1000 and 3000~eV), while the DSEP is given for 1000~eV for two different surface crossing angles of 0$^\circ$ and
70$^\circ$. It is seen that for the considered energy loss
range, being  significantly smaller than the probing energy, the shape of the DIIMFP hardly depends on the energy, or,
in other words, that the {\em normalized} distribution of energy losses is independent of the incoming energy
$w_b(T,E)\simeq w_b(T)$ to  a good approximation. The same holds for the DSEP (not shown). On the other hand, the
angular dependence of the {\em normalized}ÊDSEP is seen to be very weak as well, $w_s(T,E,\theta)\simeq w_s(T)$. Of
course the {\em total} surface excitation probability  $\langle n_s (E,\theta)  \rangle $ exhibits a pronounced angular
and energy dependence, while the energy dependence of the  IMFP, $\lambda_i(E)$,  that governs the total bulk scattering
probability, is appreciable as well.

Comparison of these quantities for Si and Au clearly reveals the pronounced difference of the electronic structure of
these materials: while the inelastic interaction in Si is governed by the collective plasmon-modes of the weakly bound
solid state
electrons, inter- and intraband transitions determine the shape of the energy loss distributions of Au. 

Another noteworthy feature in this figure is the negative excursion of the DSEP, being very pronounced for Si, while
it is much weaker for Au, but still clearly distinguishable. This is a consequence of the coupling between the bulk
and surface modes, that are orthogonal, and is commonly refered to as {\em begrenzungseffect} after the German word
for boundary\cite{ritchiepr106}. In other words, in the presence of the surface, the intensity of the volume modes is
decreased owing to the depolarization of the surface charge by the surface modes. This is clearly seen in the case of Si
where the negative excursion peaks exactly at the energy loss corresponding to a volume plasmon loss. This means that
the DSEP in fact consists of two terms: the pure surface term, which is positive, and the begrenzungs- or coupling
term, which is negative. It also makes it clear that surface and bulk excitations are  different modes
of the same phenomenon and that the distinction between these two types of inelastic scattering is essentially
artificial.  Nonetheless, for practical purposes it is useful to make this distinction in the sense that volume
excitations are considered as those loss processes that occur in an infinite boundless medium, while surface
excitations are defined as all changes in the loss probability due to the presence of a boundary. Thus, in accordance
with this definition, the DSEP is in fact a difference of two loss probabilities, the pure and the coupling term and
a negative excursion in the  DSEP is observed whenever the latter exceeds the former.

In conclusion of this section, it is noted that several semi-emipirical formulae have been published to estimate the IMFP
\cite{tansia21} and the SEP \cite{wersuxssl}. The latter quantity is given in terms of a material parameter $a_s$,  the
so--called surface excitation parameter,  by the formula:
\begin{equation}
\label{esepemp}
\langle n_s(E,\theta) \rangle =\frac{1}{a_s\sqrt{E}\cos\theta+1}
\end{equation}
The quantity $a_s$ is given for a large number of elemental solids in Ref.~\cite{wersuxssl} in units of the free electron value $a_{NFE}=\sqrt{8a_0/\pi^2e^2}=0.173
eV^{-1/2}$, where   $e^2=14.4eV\AA$ is the elementary charge squared and  $a_0=0.52\AA$ is the Bohr radius.
An empirical relationship between the surface excitation parameter $a_s$ and the generalized plasmon energy was also
derived, allowing one to estimate the extent of surface excitations for an arbitrary material.

\subsection{Multiple scattering}

The  energy and direction of motion of a charged particle travelling through a solid is changed repeatedly by multiple scattering processes. 
For non-crystalline materials, where coherent scattering can be neglected \cite{smewerjes137} the fluctuations after
multiple scattering can be expressed in terms of the fluctuation distributions in a single collision, that were
introduced in the previous section, by solving a linearized Boltzmann-type kinetic equation \cite{werepes3}.
The Green's function of this problem can be expressed in terms of the $(n-1)$--fold selfconvolution of the single
scattering fluctuation distributions weighted with the collision statistics, i.e. the number of times $n$ a given
scattering process occurs for the considered boundary conditions [Eq. (12) in Ref.~\cite{werepes3}]. The 
resulting  spectrum, or yield,  $Y(E)$ (for one type of inelastic scattering) is then found by superposition:
\begin{equation}
\label{eyield}
Y(E)=\sum\limits_{n=0}^\infty A_{n} \Gamma_{n}(T)\otimes  f_0(E+T)
\end{equation}
Here $f_0(E)$ is the energy distribution at the source and the symbol "$\otimes$" denotes a convolution over the energy
loss $T$. The quantities $\Gamma_{n}(T)$ represent the (normalized) distribution of energy losses after $n$ collisions
and are given by the $(n-1)$--fold selfconvolution of the single scattering fluctuation distribution $w(T)$ \cite{werepes3}:
\begin{eqnarray}
\label{egamma}
\Gamma_{n=0}(T)&=&\delta(T)\nonumber\\
\Gamma_{n}(T)&=&\Gamma_{n-1}(T')\otimes  w(T-T')
\end{eqnarray}
where $w(T)$ is the {\em normalized} distribution of energy losses in an individual collision.

The partial intensities $A_{n}$ represent the number of electrons that arrive in the detector after being
$n$--fold inelastically scattered in the solid  and are given by an integral over all possible lengths of the paths $s$
taken by the particle in the solid:
\begin{equation}
\label{epipld}
A_{n}=\int\limits_0^\infty  W_{n}(s) Q(s) ds
\end{equation}
Here $Q(s)$ is the distribution of pathlengths and $W_{n}(s)$ is the stochastic process for multiple scattering, which,
in the quasielastic regime, is given by \cite{werprcsd}:
\begin{equation}
\label{epoisson}
W_{n}(s)=\Big(\frac{s}{\lambda_i}\Big)^{n}\frac{e^{-s/\lambda_i}}{n!}
\end{equation}

In a REELS experiment,  the incoming electron experiences surface excitations on the way into the solid, is multiple
scattered in the bulk and again suffers a surface electronic  energy loss when crossing the solid--vacuum boundary on its
way out of the solid. In the following, the ingoing and outgoing part of the trajectory will be denoted by the
superscripts "i" and "o", respectively. When interference effects between the different types of inelastic scattering can
be neglected \cite{werqsasia}, the total energy fluctuation  distribution for all types of scattering is given by a convolution of the
individual fluctuation distributions   and  Equation~(\ref{eyield}) may be generalized as follows: 
\begin{equation}
\label{eyieldreels}
Y(E)=
\sum\limits_{n_b=0}^\infty
\sum\limits_{n_s^i=0}^\infty
\sum\limits_{n_s^o=0}^\infty
 A_{n_b,n_s^i,n_s^o}
\Gamma_{n_b}(T)\otimes\Gamma_{n_s ^i}(T')\otimes\Gamma_{n_s^o}(T'') \otimes f_0(E+T+T'+T'')
\end{equation}

As mentioned before, the width of the surface
scattering zone $v/\omega_s$ is smaller than, or of the order of, the elastic mean free path $\lambda_e$. 
In this case the  electrons path in the surface scattering zone is rectilinear to a good approximation.
A most beneficial consequence of this fact is that the partial intensities for the three types of scattering are uncorrelated \cite{wernrelprl}:
\begin{equation}
\label{epiuncorr}
A_{n_b, n_s^i, n_s^o}=  A_{n_b}\times A_{n_s^i}\times A_{n_s^o}
\end{equation}
Although (small) effects of  deflections in the surface scattering zone have been experimentally observed for a rather
pathological case \cite{werzemsep}, implying that Eqns.~(\ref{epiuncorr}) (and (\ref{episurf}) below) are not strictly true,
this approach nonetheless has proven to constitute an effective approximation \cite{werzemsep,wernrelprl,wernrelsia,wercoissl}.

Since there exists  a  unique straight line path connecting any two points in space, the part of the pathlength
distribution relevant for surface excitations  resembles a $\delta$--function, $Q_s(s)\approx  \delta(s-v/\omega_s\mu_i)$
when the 
passage through the surface scattering zone is approximately rectilinear.
 In this case Equation~(\ref{epipld}) can readily  be integrated giving:
\begin{equation}
\label{episurf}
 A_{n_s^i}(\mu_i)=\frac{\langle n_s^i(\mu_i)\rangle^{n_s^i}}{n_s ^i!}e^{-\langle n_s^i(\mu_i)\rangle}
\end{equation}
Where the shorthand notation $\mu_i=\cos\theta_i$ was used to indicate the incident polar direction of surface crossing.
This reveals that the average number of surface excitations may alternatively be expressed as:
\begin{equation}
\label{esep}
\langle n_s^i (\mu_i) \rangle\approx  \frac{v}{\omega_s\lambda_i^{s,i}\mu_i},
\end{equation}
where $\lambda_i^{s,i}$ is the effective inelastic mean free path for surface scattering along the incoming part of the
trajectory. Analoguous expressions hold for the outgoing partial intensities.

For bulk inelastic scattering the situation is more complicated.
The  distribution of  pathlengths depends in a complex manner on the incident and outgoing angle and the energy of the
particle  \cite{werhayreels,werzemsep,werqsasia,wernrelsia}. Therefore the partial intensities for bulk scattering,
$A_{n_b}$, are most conveniently established by calculating the pathlength distribution by some
numerical procedure, e.g., a Monte Carlo calculation \cite{werqsasia,werepesjes}, and using Equation~(\ref{epipld}).

Examples for the bulk partial intensities are given in Figure~\ref{fcn} for electrons of various energies reflected
from a Si and a Au surface for normal incidence and for an off-normal emission angle of 60$^\circ$, corresponding to
the geometrical configuration used throughout this work. The sequence of partial intensities is seen to be
qualitatively different for the two considered materials and also for the different energies considered. This is a
well known effect \cite{werhayreels,werepesjes,werzemsep} that is caused by the relative strength of elastic and
inelastic scattering for a given geometrical configuration, being governed by the complex energy and angular
dependence of the elastic scattering cross section.

To simplify Equation~(\ref{eyieldreels}), the reduced spectrum $y(E)$ is introduced. This is the spectrum 
divided by the area of the elastic peak. The latter is per definition given by
the zero  order partial intensity  $A_{n_b=0,n_s^i=0,n_s^o=0}$. It will furthermore be assumed that the width of energy
distribution at the source (i.e. the thermal spread in the electron gun) is negligible compared to the width of any
feature  in the relevant differential mean free paths, implying that the source energy distribution can be replaced by a
$\delta$--function,
$f_0(E)=\delta(E-E_0)$ where $E_0$ is the source energy. For the following manipulations, it is convenient to consider
the Fourier transform of the spectrum, since, by virtue of the convolution theorem, this quantity can be written as a
power series in the respective mean free paths [cf. Equation~(\ref{egamma})]. Below, any quantity in Fourier space will
be indicated by a tilde sign ("$\widetilde{~}$"). The Fourier transform of the  reduced spectrum is given by:
\begin{equation}
\label{eyieldtmp}
\widetilde{y}=
\sum\limits_{n_b=0}^\infty
\sum\limits_{n_s^i=0}^\infty
\sum\limits_{n_s^o=0}^{n_s^i}
 \alpha_{n_b,n_s^o, n_s^i-n_s^o}
\widetilde{w}_b^{n_b}
\widetilde{w}_{s,i}^{n_s^o}
\widetilde{w}_{s,o}^{n_s^i-n_s^o}
\end{equation}
where the reduced partial intensities $\alpha_{n_b,n_s^i ,n_s^o}=A_{n_b, n_s^i, n_s^o} /A_{n_b=0, n_s^i=0, n_s^o=0}$
were introduced. Note that the order of the summation   in Equation~(\ref{eyieldtmp}) was also changed. Using
expression (\ref{epiuncorr}) and (\ref{episurf}) and the binomial theorem, one finds:
\begin{equation}
\label{eyieldtmp2}
\widetilde{y}=
\sum\limits_{n_b=0}^\infty
\alpha_{n_b}
\widetilde{w}_b^{n_b}
\sum\limits_{n_s=0}^\infty
\frac{
(\langle n_s^i\rangle\widetilde{w}_{s,i}+
\langle n_s^o\rangle\widetilde{w}_{s,o})^{n_s}
}{n_s!}
\end{equation}
where $n_s=n_s^i+n_s^o$.
Since the results in Figure~\ref{fdiimfp} clearly show that the normalized DSEP only depends  weakly on the surface
crossing angle, it is reasonable to put $w_{s,i}(T)\approx w_{s,o}(T)\equiv w_s(T)$,  in this way combining the effects
of  surface excitations along the in- and outgoing part of the trajectory. Going back to real space, this gives:
\begin{equation}
\label{eyieldfinal}
y(E)=
\sum\limits_{n_b=0}^\infty
\sum\limits_{n_s=0}^\infty
\alpha_{n_b}
\alpha_{n_s}
\Gamma_{n_b}(T)\otimes
\Gamma_{n_s}(T')\otimes \delta(E-E_0+T+T')
\end{equation}
where $\alpha_{n_s}$ are the reduced partial intensities for surface scattering 
\begin{equation}
\alpha_{n_s}=\frac{A_{n_s}}{A_{n_s=0}}=\frac{\langle n_s \rangle^{n_s}}{n_s!}
\end{equation}
 and with $ \langle n_s\rangle=\langle n_s^i(\mu_i)\rangle+\langle n_s^o(\mu_o)\rangle$. 

Finally, the elastic peak is removed from the spectrum and the energy scale is converted to an energy loss scale giving the reduced {\em loss} spectrum $y_L(T)$ as:
\begin{equation}
\label{eyieldloss}
y_L(T)=
\sum\limits_{n_b=0}^\infty
\sum\limits_{n_s=0}^\infty
\alpha_{n_b, n_s}
\Gamma_{n_b}(T')
\otimes
\Gamma_{n_s}(T-T')
\end{equation}
with $\alpha_{n_b=0, n_s=0}=0$. This form of the loss spectrum is used in the further analysis.

\subsection{Decomposition of REELS spectra}
It is the objective of the present work to extract the DIIMFP and DSEP from experimental loss spectra. Recognizing
Equation~(\ref{eyieldloss}) as a bivariate power series (in $\widetilde{w}_b$ and $\widetilde{w}_s$) in Fourier
space, it is immediately obvious that this equation has no unique solution. However, when {\em two } loss spectra with
different partial intensities 
\begin{eqnarray}
\label{etwospectra}
\widetilde{y}_{L,1}
&=&
\sum\limits_{n_s=0}^\infty
\sum\limits_{n_b=0}^\infty
\alpha_{n_s , n_b} 
\widetilde{w}_b^{n_b}
\widetilde{w}_s^{n_s}
\nonumber\\
\widetilde{y}_{L,2}
&=&
\sum\limits_{n_s=0}^\infty
\sum\limits_{n_b=0}^\infty
\beta_{n_s , n_b} 
\widetilde{w}_b^{n_b}
\widetilde{w}_s^{n_s},
\end{eqnarray}
with $\alpha_{0,0}=\beta_{0,0}=0$, are measured, reversion of the bivariate power series becomes possible.
Experimentally, this implies that two loss spectra need to be acquired at different energies and/or geometrical
configurations. From the theoretical point of view, we need to make the assumption, in this case, that the normalized DIIMFP
and DSEP are independent of the surface crossing angle and/or the energy to a good approximation (cf.
Figure~\ref{fdiimfp}). 

Formally, the reversion of this bivariate power series is effected by the  expansion:
\begin{eqnarray}
\label{eformalreversion}
\widetilde{w}_b
&=&
\sum\limits_{p=0}^\infty
\sum\limits_{q=0}^\infty
u_{p,q}^b 
\widetilde{y}_{L,1}^{p}
\widetilde{y}_{L,2}^{q}
\nonumber\\
\widetilde{w}_s
&=&
\sum\limits_{p=0}^\infty
\sum\limits_{q=0}^\infty
u_{p,q}^s 
\widetilde{y}_{L,1}^{p}
\widetilde{y}_{L,2}^{q}
\end{eqnarray}
with $u_{0,0}^b=u^s_{0,0}=0$.
This can be seen by substituting  Equation~(\ref{etwospectra}) back into  Equation~(\ref{eformalreversion}) and equating
coefficients of equal powers of $\widetilde{w}_b$ and $\widetilde{w}_s$. This  gives the equations for the unknown
coefficients $u_{p,q}^b$ and $u^s_{p,q}$. 

In doing so, one is faced with the problem of evaluating the $p$-th power of
$\widetilde{y}_{L,1}$ times the $q$-th power of
$\widetilde{y}_{L,2}$ that can be expressed as:
\begin{equation}
\label{etensorgamma}
\widetilde{y}_{L,1}^p
\widetilde{y}_{L,2}^q=
\sum\limits_{n_s=0}^\infty
\sum\limits_{n_b=0}^\infty
\gamma_{p,q,n_b,n_s}
\widetilde{w}_b^{n_b}
\widetilde{w}_s^{n_s}
\end{equation}
The components  of the tensor $\gamma_{p,q,n_b,n_s}$  are found to be  given by: 
\begin{equation}
\label{egammacomponents}
\gamma_{p,q,n_b,n_s}=
(\alpha_p,\beta_q)_{(n_b,n_s)}^{(p+q)}
\end{equation}
These components are equal to the sum of all possible terms with $p$ factors in $\alpha_{k,l}$ and  $q$ factors in
$\beta_{m,n}$, whose indices "add up" to the target index combination $(n_s,n_b)=(k+m,l+n)$. For example:
\begin{eqnarray}
\label{egammaexamples}
(\alpha_2)_{(2,1)}^{2}&=&
\alpha_{1,0}\alpha_{1,1}+
\alpha_{0.1}\alpha_{2,0}+
\alpha_{1,1}\alpha_{1,0}+
\alpha_{2,0}\alpha_{0,1}
\nonumber\\
(\alpha_1,\beta_2)_{(1,2)}^{3}&=&
\alpha_{0,1}\beta_{0,1}\beta_{1,0}+
\alpha_{0,1}\beta_{1,0}\beta_{0,1}+
\alpha_{1,0}\beta_{0,1}\beta_{0,1}
\end{eqnarray}
Obviously,  one has
\begin{equation}
\label{egammais0} 
\gamma_{p,q,n_b,n_s}=0~~~~~~~~~~~~for~all~p+q>n_s+n_b
\end{equation}
since the target index combination $(n_s,n_b)$ can only be
produced by a number of factors less than or equal to $(n_s+n_b)$ when $\alpha_{0,0}=\beta_{0,0}=0$ [see
Equation~(\ref{etwospectra})]. Furthermore, one has:
\begin{eqnarray}
\label{ethis}
\gamma_{n_b,n_s,1,0}&=&\alpha_{n_b,n_s}
\nonumber\\
\gamma_{n_b,n_s,0,1}&=&\beta_{n_b,n_s}
\end{eqnarray}
Based on these guidelines, the tensor $\gamma_{p,q,n_b,n_s}$ can
readily be established by means of a recursive algorithm for any sequence of partial intensities $\alpha_{n_b, n_s} $ and
$\beta_{n_b , n_s} $.

Inserting Equation~(\ref{etwospectra}) into Equation~(\ref{eformalreversion}) and equating coefficients, one
finds a set of  two equations with two unknowns  for the first order bulk coefficients:
\begin{eqnarray}
\label{eupqbeq}
1&=&
\sum\limits_{p=0}^\infty
\sum\limits_{q=0}^\infty
u^b_{p,q}\gamma_{p,q,1,0}
=u_{0,1}\beta_{1,0}+u_{1,0}\alpha_{1,0}
\nonumber\\
0&=&
\sum\limits_{p=0}^\infty
\sum\limits_{q=0}^\infty
u^b_{p,q}\gamma_{p,q,0,1}=
u_{0,1}\beta_{0,1}+u_{1,0}\alpha_{0,1}
\end{eqnarray}
The solution is:
\begin{eqnarray}
\label{eupqb1}
u_{1,0}^b&=&\frac{\beta_{0,1}}{\beta_{0,1}\alpha_{1,0}-\beta_{1,0}\alpha_{0,1}}
\nonumber\\
u_{0,1}^b&=&\frac{\alpha_{0,1}}{\alpha_{0,1}\beta_{1,0}-\alpha_{1,0}\beta_{0,1}}
\end{eqnarray}
Similarly, for the first order surface coefficients one finds:
\begin{eqnarray}
\label{eupqs1}
u_{1,0}^s&=&\frac{\beta_{1,0}}{\beta_{1,0}\alpha_{0,1}-\beta_{0,1}\alpha_{1,0}}
\nonumber\\
u_{0,1}^s&=&\frac{\alpha_{1,0}}{\alpha_{1,0}\beta_{0,1}-\alpha_{0,1}\beta_{1,0}}
\end{eqnarray}
The equation determining the higher order ($p+q>1)$ coefficients
\begin{equation}
\label{eupqhigh1}
0=
\sum\limits_{p=0}^\infty
\sum\limits_{q=0}^\infty
u_{p,q}
\gamma_{p,q,n_b,n_s}
\end{equation}
can be split into two parts by using property (\ref{egammais0}),  changing the order of the summation and writing:
\begin{equation}
\label{eupqhigh2}
0=
\sum\limits_{p=0}^{n_s+n_b-1}
\sum\limits_{q=0}^p
u_{q,p-q}
\gamma_{q,p-q,n_b, n_s}
+
\sum\limits_{p=0}^{n_s+n_b}
u_{p,n_s+n_b-p}
\gamma_{p,n_b+n_s-p,n_b,n_s}
\end{equation}
The first term represents a $(n_s+n_b+1)$--dimensional vector containing the coefficients $u_{p<n_b, q<n_s}$, which
have been established during the previous step of the algorithm.
The latter term is the product of a square matrix with the same dimension and the unknown vector $(u_{0,n_b+n_s},
u_{1,n_b+n_s-1},  u_{2,n_b+n_s-2}, ...u_{n_b+n_s,0})$. Thus, for each value of $n_b+n_s$, the corresponding coefficients
are obtained by solution of a system of $n_b+n_s+1$ linear equations with as many unknowns. Consecutively performing this procedure for
values of the total scattering order $n_b+n_s=1,2,....n_{max}$, where $n_{max}$ is the collision order where convergence
of the series Equation~(\ref{eformalreversion})  is attained, leads to the desired coefficient matrices $u_{p,q}^b$ and
$u_{p,q}^s$. Note that Eqns.~(\ref{eupqhigh1}) and (\ref{eupqhigh2}) are identical for the surface and bulk coefficients,
only the calculation for the first order term is different [see Eqns.~(\ref{eupqb1}) and (\ref{eupqs1})]

Finally, having established the bulk and surface expansion coefficients,  the Fourier backtransform of
Equation~(\ref{eformalreversion}) gives:
\begin{eqnarray}
\label{ereversionrealspace}
{w}_b (T)
&=&
\sum\limits_{p=0}^\infty
\sum\limits_{q=0}^\infty
u_{p,q}^b 
{y}_{L,1}^{(p)} (T')\otimes
{y}_{L,2}^{(q)}(T-T')
\nonumber\\
{w}_s(T)
&=&
\sum\limits_{p=0}^\infty
\sum\limits_{q=0}^\infty
u_{p,q}^s 
{y}_{L,1}^{(p)}(T')\otimes
{y}_{L,2}^{(q)}(T-T')
\end{eqnarray}
where ${y}_{L,1}^{(p)}$ and ${y}_{L,2}^{(q)}$ denote the $(p-1)$--fold selfconvolution of $y_{L,1}$ and the $(q-1)$-fold
selfconvolution of $y_{L,2}$, respectively.

\section{Results and Discussion}
The procedure outlined above was implemented and applied to simulated data, to test its performance in terms of
numerical stability and accuracy. Subsequently it was applied  to experimental REELS spectra, to gain information on
the dielectric response of several solids to incoming electrons. In all cases, the pair of spectra
considered was a set of two REELS spectra taken at different energies, but for the same geometrical configuration.

Calculation of the tensor $\gamma_{p,q,n_b,n_s}$ was performed using a simple recursive algorithm (about 20 lines of
code), following the guidelines given in the previous section. While computation of the first few orders ($p+q\le 7$)
takes several seconds on a modern PC, the 8-th order alone takes 10 seconds, the 9th order 10 ten minutes, the 10th
order takes several hours and the 11th order several weeks. Although faster algorithms for reversion of multivariate
power series are available \cite{raney,brent,cheng}, these are quite complex and were not considered since convergence of
the proposed algorithm is typically attained for $n_{max}\le7$ for energy loss ranges extending up to 100~eV. 

The convergence of the algorithm can be assessed in a simple way by using the retrieved DSEP and DIIMFP to simulate
the loss spectra and compare these with the original loss spectra from which these quantities were retrieved. For
the energy range for which the procedure has attained convergence, these simulated spectra are {\em exactly} identical to
the input spectra, while for higher energies, corresponding to scattering orders beyond the value of $n_{max}$, they
diverge rapidly. In this way the required value of $n_{max}$ can easily be determined for a specific application. For the
results shown in the present work, the maximum energy loss considered was 100~eV and the procedure converged for
$n_{max}=5-7$.

\begin{figure}[htb]
\ifpdf
{\includegraphics[width=17.0cm]{fcn.pdf}}
\else
\epsfxsize = 0.8\hsize
\epsffile{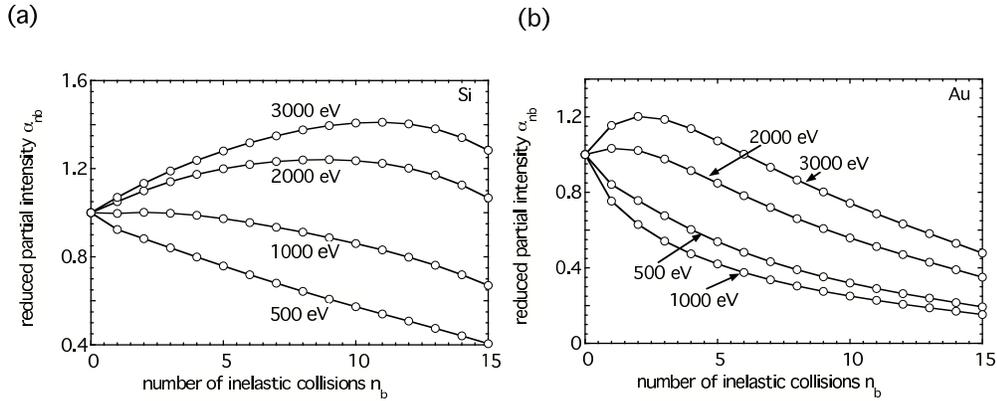}
\fi
\caption{%
Reduced partial intensities for quasi--elastic reflection,  $\alpha_{n_b}=A_{n_b}/A_{n_b=0}$ for volume scattering for
several energies. These model calculations were performed for normal incidence and for an off-normal emission
direction of 60$^\circ$(a.)
Si; (b.) Au.}
\label{fcn}
\end{figure}

It was found that the outlined algorithm is very robust when a reasonable choice for the difference of the 
partial intensities of the two input spectra is made.  The stability of the procedure is governed by the quantity
$\Delta=| \beta_{1,0}\alpha_{0,1}-\beta_{0,1}\alpha_{1,0}|$ which should not be too much smaller than unity
$\Delta\agt 0.1$ when the procedure is applied to experimental data.  On the other hand the difference in energy/emission
angle should not be too large since then the  difference of the normalized DIIMFP/DSEP for the different experimental
conditions becomes more pronounced. This implies that one should select that
combination of energies for which this criterion is optimally fullfilled (see Figure~\ref{fcn}).

Spectra were simulated by calculation of the pathlength distribution using a Monte Carlo algorithm
\cite{werepesjes} and using Equation~(\ref{epipld}) to calculate the bulk partial intensities. The inelastic mean free
path was derived from the TPP-2M formula \cite{tansia21}, and elastic cross sections were calculated with the
computer code of Ref.\cite{yatcpc} for a Thomas--Fermi--Dirac potential \cite{bonchem}. The surface partial intensities
were obtained from Equation~(\ref{episurf}), using the values for the SEP given in Ref.~\cite{wersuxssl}. The DSEP
and DIIMFP were  established by means of Eqns.~(\ref{ediimfp}) and (\ref{edsep}) using the Drude-Lindhard parameters for
the dielectric function in Ref.~\cite{tungpr49}.

Three different types of simulated spectra were analyzed before the procedure was applied to experimental data, in
order to assess the influence of several assumptions that are made in this procedure on the outcome of a retrieval
operation: (I.)  loss spectra that were simulated for an elastic peak modelled by a true $\delta$-function
and using identical normalized DIIMFPs and DSEP for both energies/geometries of the spectrum pair; (II.) loss spectra that were
simulated  taking into account the energy/angular dependence of the normalized loss distributions using an infinitely
sharp elastic peak; and (III.) loss spectra with different DIIMFP and DSEP at the considered energies/geometries and using an
elastic peak with a finite width. These different types of model calculations will be refered to below by their roman
numeral indicated above.

\begin{figure}[htb]
\ifpdf
{\includegraphics[width=17.0cm]{fdiimfp.pdf}}
\else
\epsfxsize = 0.8\hsize
\epsffile{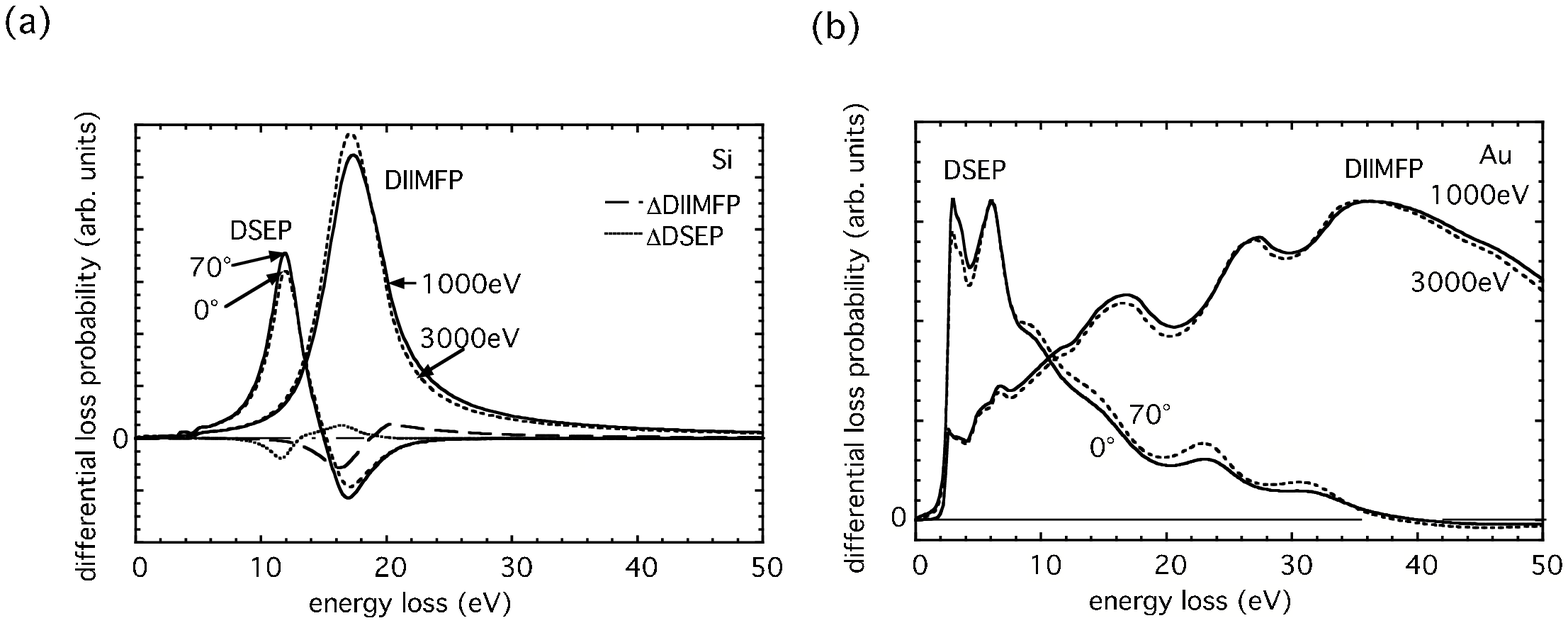}
\fi
\caption{%
Differential inverse inelastic mean free path (DIIMFP) and differential surface excitation probability (DSEP) for
medium energy electrons in Si and Au. The DIIMFP is presented for 1000~eV(solid curves) and 3000~eV (dashed curves)
while the DSEP is shown for 1000~eV for two different angles of surface crossing (0$^\circ$, solid curves and
70$^\circ$, dashed curves).  The DSEP was divided by a factor of 5 to facilitate comparison. The curves labelled
$\Delta$DSEP and $\Delta$DIIMFP  in (a) represent the difference of the DSEP and DIIMFP for the considered  angle of
surface crossing and energy respectively.
(a.) Si; (b.) Au.}
\label{fdiimfp}
\end{figure}

The type I model calculations always returned the exact model DIIMFP and
DSEP (within the numerical precision of the PC), even when a significant amount of noise was added to the model spectra.
When the energy dependence of the DIIMFP and DSEP is accounted for in the simulation (type II model calculations), minor
deviations were observed, of the order of the deviations seen for different energies/angles shown  in Figure~\ref{fdiimfp}.
These deviations were more pronounced --but still minor--  for sharply peaked loss distributions (e.g., for Si). When a
finite width of the elastic peak was used (type III model calculations), the retrieved loss distributions become
smoother. Results of type III model calculations  are shown and discussed below where the  retrieval results using
experimental data are presented.

The procedure used to acquire the experimental data  has been described in detail before \cite{werepesjes}. REELS data
were taken for  20 elemental solids  (Ag, Al, Au, Be, Bi, C, Co, Cu, Fe, Ge, Mo, Ni, Pd, Pt, Si, Ta, Te, Ti, V, W) in the
 energy range between 300 and 3400~eV for normal incidence and an off-normal emission angle of 60$^\circ$, using a
hemispherical analyzer operated in the constant analyzer energy mode giving a width of the elastic peak of 0.7~eV. Count
rates in the elastic peaks were kept well below the saturation count rate of the channeltrons  and a dead time
correction was applied to the data. For each material the optimimum energy combination for the retrieval procedure of 
two loss spectra was determined by inspection of the partial intensities. For most materials the  optimum energy
combination was (1000-3000~eV).

\begin{figure}[htb]
\ifpdf
{\includegraphics[width=12.0cm]{flosssi.pdf}}
\else
\epsfxsize = 0.8\hsize
\epsffile{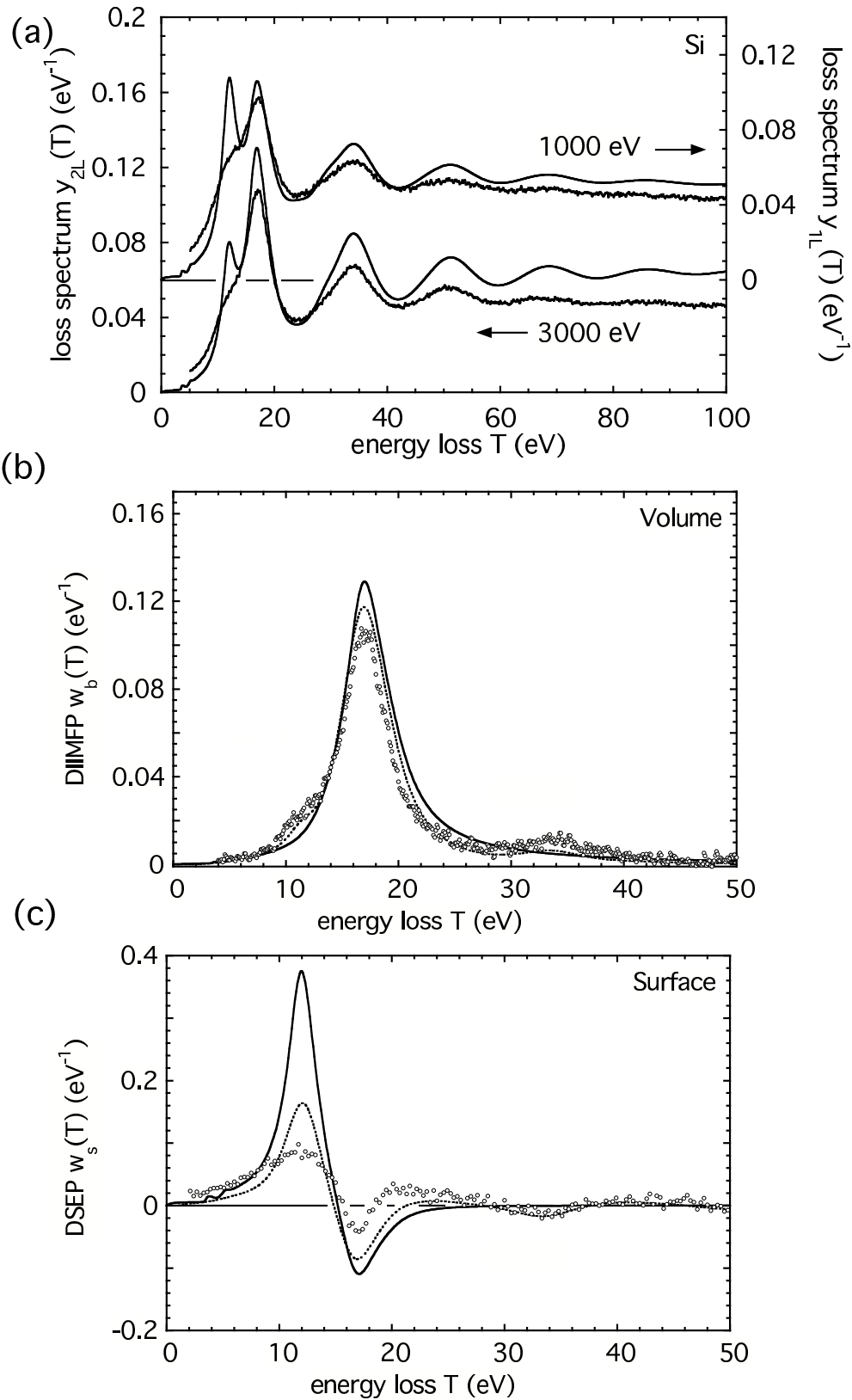}
\fi
\caption{%
(a.) Measured energy loss spectra for 1000 and 3000~eV for Si after removal of the elastic peak  (noisy curves). The
smooth curves are the simulated loss spectra using theoretical shapes for the DSEP and DIIMFP. 
(b.) Retrieved normalized DIIMFP (open circles). The solid curve is the theoretical result Equation~(\ref{edsep}).  The
dashed curve represents  the retrieved volume loss distribution using the simulated spectra in (a.) as input.  (c.) same
as (b) for the DSEP. (see text)}
\label{flosssi}
\end{figure}

\begin{figure}[htb]
\ifpdf
{\includegraphics[width=14.0cm]{flosscu.pdf}}
\else
\epsfxsize = 0.8\hsize
\epsffile{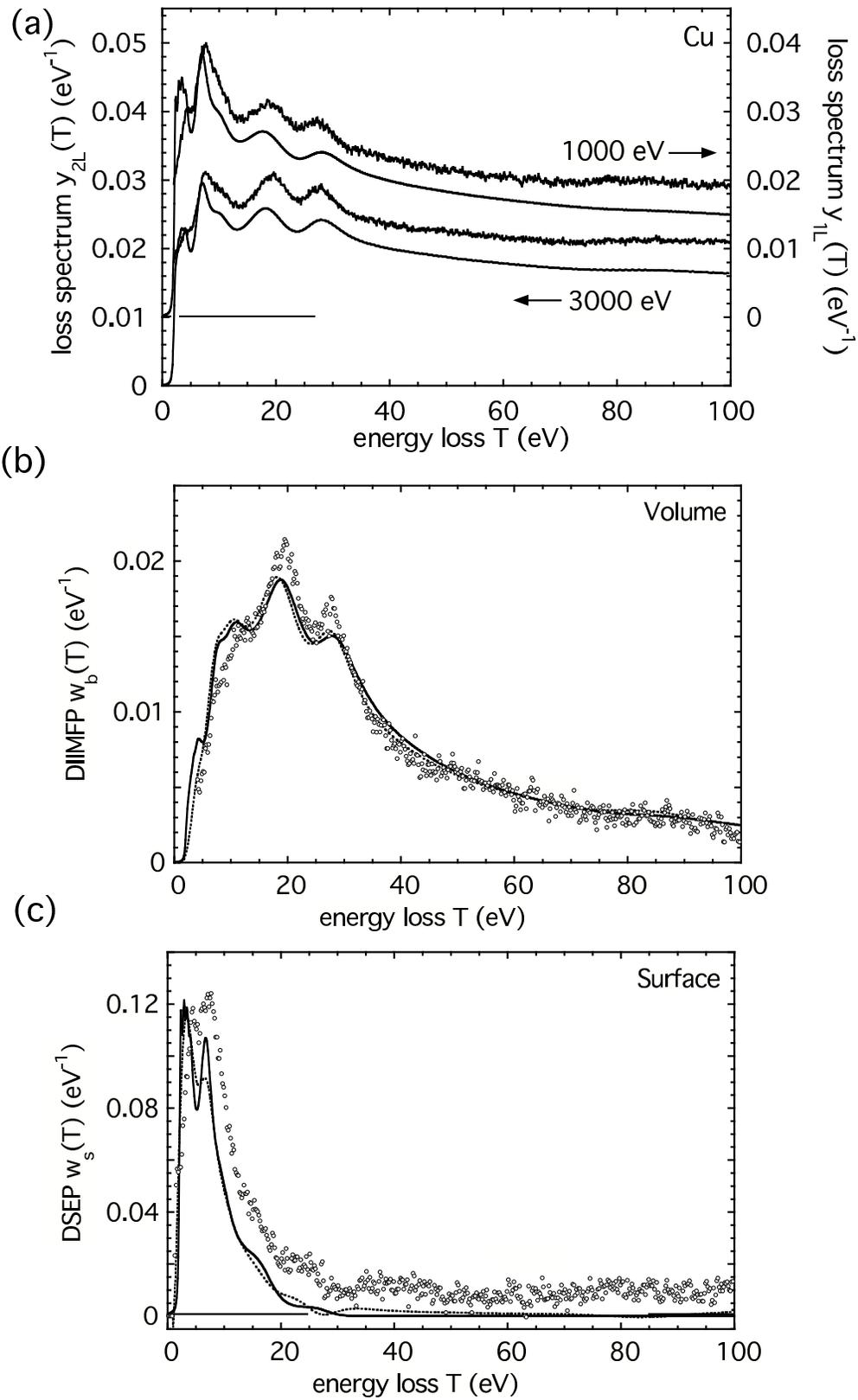}
\fi
\caption{%
Same as Figure~\ref{flosssi} for Cu.
}
\label{flosscu}
\end{figure}

\begin{figure}[htb]
\ifpdf
{\includegraphics[width=14.0cm]{flossau.pdf}}
\else
\epsfxsize = 0.8\hsize
\epsffile{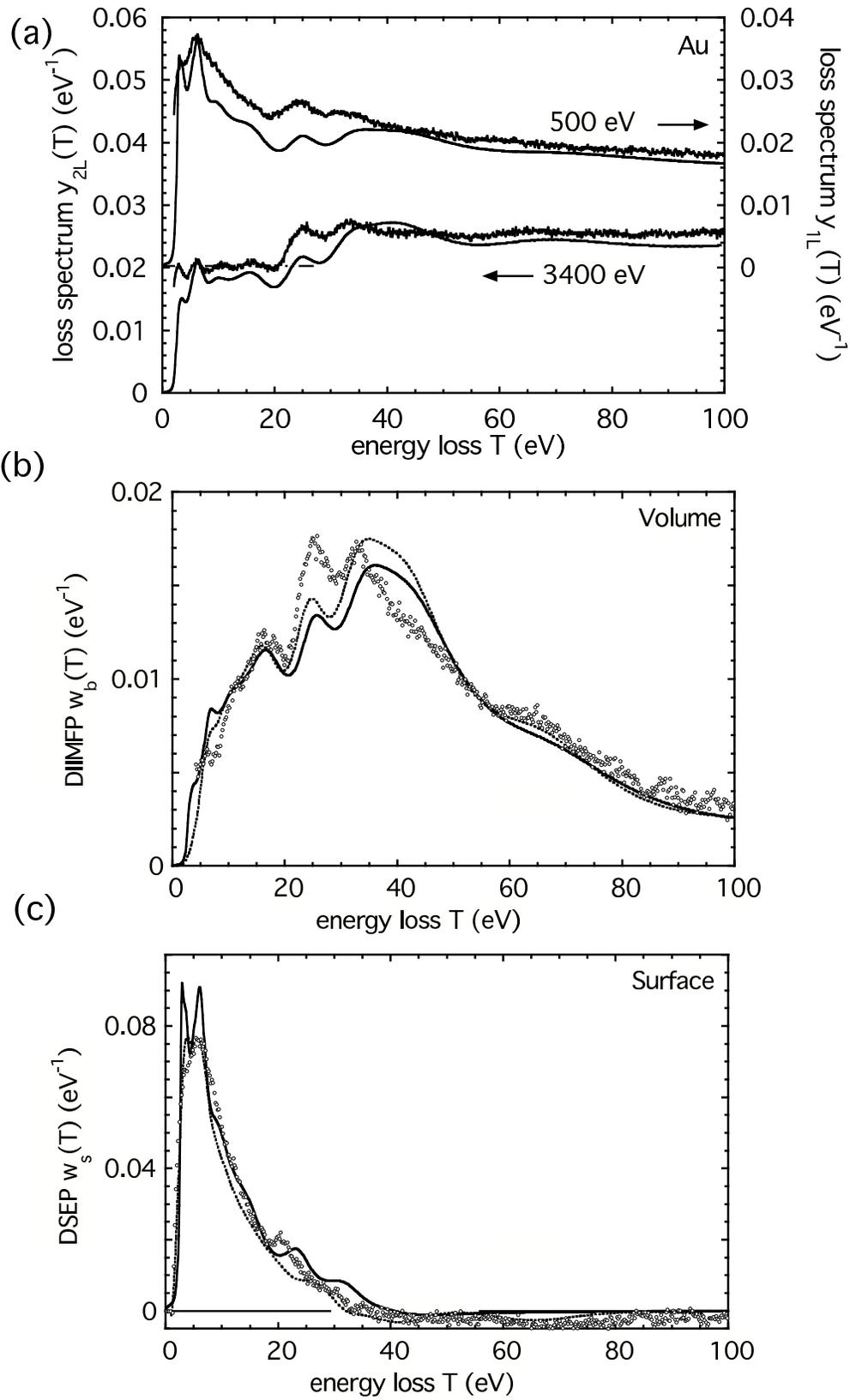}
\fi
\caption{%
Same as Figure~\ref{flosssi} for Au for measured spectra taken at 500 and 3400~eV.
}
\label{flossau}
\end{figure}

Figure~\ref{flosssi}--\ref{flossau} show the results for Si, Cu and Au. 
In (a.) the experimental loss spectra used as input are represented by the noisy curves. These loss spectra were obtained
from the REELS spectra by fitting the elastic peak to a Gaussian. Subsequently, the experimental data were divided by the
area under this curve, the fitted peak was subtracted from them and
the energy scale was converted to an energy loss scale. Finally, the measured spectrum $S_L(T)$ [in counts per channel]
was converted to experimental yield $y_L(T)$ [in reciprocal eV],  corresponding to Equation~(\ref{eyieldloss}), by division by the
channel width $\Delta E$.

The smooth curves in (a.) are the simulated spectra for an elastic peak width matching that of the experimental data and
with the energy dependence of the normalized loss distributions taken into account (type III model calculations). In
(b.) the retrieved normalized DIIMFP (open circles) is compared with the theoretical DIIMFP [Equation~(\ref{ediimfp}),
solid curves] and with the DIIMFP retrieved from the simulated spectra (type III model calculations, dotted curve). In
(c.) the retrieved normalized DSEP is compared with theory [Equation~(\ref{edsep}), solid curves] and the result based
on simulated spectra (type III model calculations, dotted curves). The results in (b.) and (c.) are presented as 
returned by the decomposition algorithm, and were not scaled in any way. Finally, the DIIMFP and DSEP retrieved from the
experimental data were used to calculate another set of model spectra which are compared in (a.) with the experimental
data, but are impossible to distinguish from the latter since these model data {\em exactly } coincide with the 
experimental data since Equation~(\ref{ereversionrealspace}) and Equation~(\ref{etwospectra}) are each others {\em
exact} inverse for the energy loss range where convergence has been attained.

The raw loss spectra for Si  shown in Figure~\ref{flosssi}a are seen to be quite similar for 1000 and 3000~eV. Close
inspection of the loss features for the fourth and fifth order plasmon reveals that the data for 1000~eV decrease
slightly faster with increasing energy loss than for 3000~eV, in full accordance with the partial intensities for
these energies, shown in Figure~\ref{fcn}a. However, the main difference between the two spectra is the relative
intensity of the first surface plasmon, being clearly lower for 3000~eV. The main difference between the experimental
data and the simulated REELS spectra is the fact that the experimental plasmon peaks are broader than the simulated
ones, and in consequence the intensity at higher loss energies is lower than for the simulated data.

The retrieved bulk loss distribution (open circles in Figure~\ref{flosssi}b) agrees fairly well with the result
predicted by theory (solid curve), but significant deviations are observed as a shoulder at around 12~eV and a spurious
peak at about twice the volume plasmon energy. However, these features are also present in the loss distribution
retrieved from the simulated spectra (type III model calculations, dotted lines).  
Furthermore, the feature at $\sim$12~eV exactly coincides with the peak in the difference distribution $\Delta$DSEP in
Figure~\ref{fdiimfp}a. The  $\Delta$DIIMFP difference distribution has a peak in the vicinity of bulk plasmon energy
implying that the theoretical sequence of partial intensities no longer matches those required for the retrieval (since
the area under the normalized DIIMFP is different). In consequence, the elimination of the second bulk plasmon is not
complete, exactly as observed in the data at about 32~eV.
 It can therefore be concluded that the deviations between the theoretical loss distributions --for Si, with quite
sharple peaked loss features-- and those retrieved from experimental data are mainly a consequence of the energy dependence of
the {\em shape} of the DIIMFP and the angular dependence of the {\em shape}Êof DSEP (see Figure~\ref{fdiimfp}), which
is ignored by Equation~(\ref{ereversionrealspace}).

For the differential surface excitation probability, a similar behaviour is seen: near 32~eV, a second negative
excursion, apart from the theoretically expected {\em begrenzungs}-effect at $\sim$16~eV, can be seen. Again this
feature is also present in the type III model data and can thus  be attributed to the energy dependence of the shape
of the DIIMFP and DSEP.
The {\em begrenzungs}-effect itself is quite nicely reproduced by the experimental loss distributions. The
only deviation from theory that cannot be explained by the assumption that the shape of the DIIMFP and DSEP do not
depend on the energy and surface crossing angle are the features below
$\sim$10~eV, being significantly higher than the type III model calculations and rendering the overall shape of the
DSEP broader than theoretically expected. This explains
the differences between the model data and experiment in Figure~\ref{flosssi}a. Such deviations were observed before
\cite{wernrelprl} but a clear explanation has not been proposed. We note, however, that the physical model for
electron reflection discussed in the present paper completely disregards direct creation of electron-hole pairs as a
result of the impact of the primary electrons on the solid.

The differences between the simulated and experimental spectra for Cu shown in Figure~\ref{flosscu} are similar to
those for Si in that the intensity of the  simulated data differs from the experimental ones and that the surface loss
features are more clearly more pronounced for 1000~eV. The width of the loss features agrees quite well in
this case. Also, in the experimental data, a faint trace of the Cu~MII edge at $\sim$77~eV is discernable which is absent
in the simulated data, being attributable to the fact that core-edges are difficult to describe by means of a
Drude-Lindhard fit to optical data. The retrieved bulk loss distributions agree satisfactorily with theory, while the
experimental surface loss distribution is slightly broader and tails off unexpectedly slowly towards higher loss
energies. The larger width of the retrieved DSEP explains the higher intensity of the experimental raw data compared
to the model calculations in Figure~\ref{flosscu}a, but the remaining intensity at loss energies above $\sim$30~eV is
not clear. It may be attributable to the uncertainty of the value of $\langle n_s\rangle$ used in the retrieval.

The results for Au are shown in Figure~\ref{flossau}.  Note that the shape of the raw data far away
from the surface loss features is distinctly different for the two energies, as predicted by the
corrresponding sequences of bulk partial intensities for these energies, shown in Figure~\ref{fcn}b.
The retrieved loss distributions again agree quite well with theory except for deviations in the DIIMFP at around
$\sim$30~eV. 

\begin{figure}[htb]
\ifpdf
{\includegraphics[width=17.0cm]{fcomp.pdf}}
\else
\epsfxsize = 0.8\hsize
\epsffile{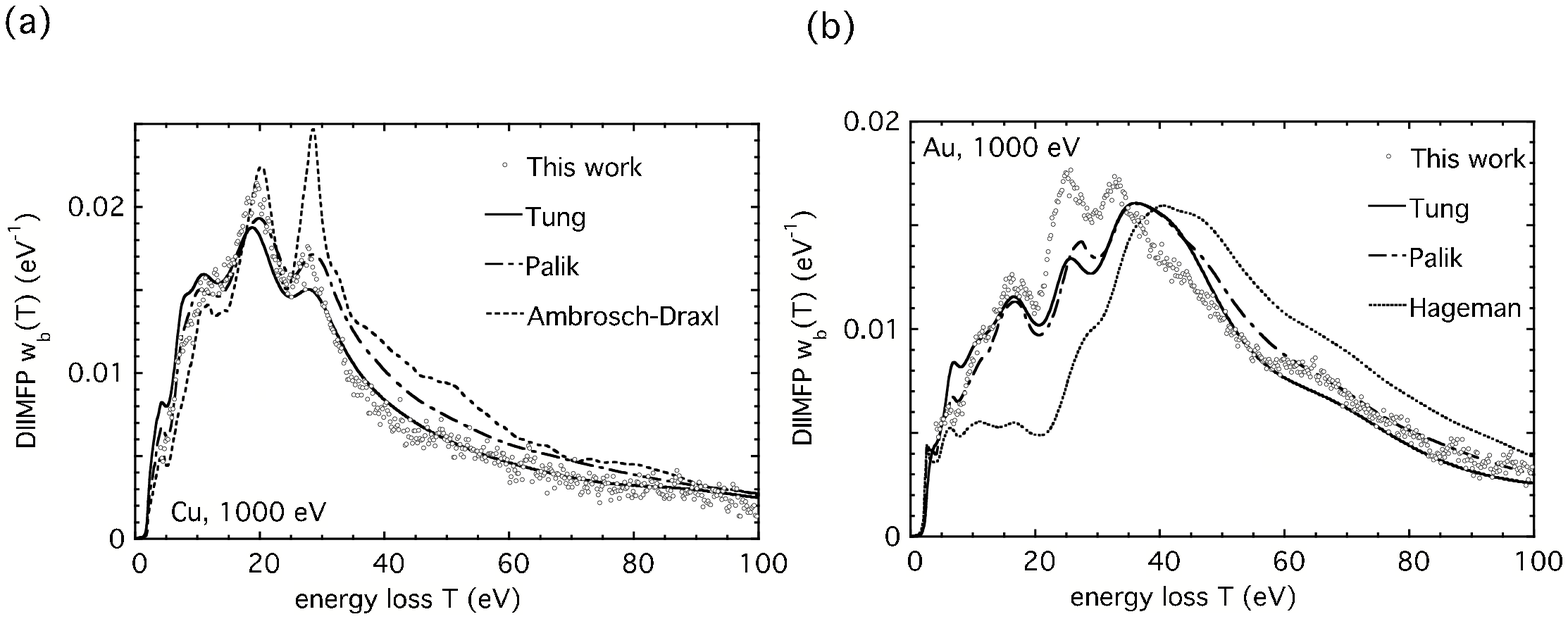}
\fi
\caption{%
Comparison of the DIIMFP obtained in the present work (open circles)  with results based on Equation~(\ref{ediimfp})
using optical data taken from different sources. 
a.) Cu, 1000 eV. Solid curve: Tung, Ref.~\cite{tungpr49};
Dash dotted curve: Palik, Ref.~\cite{palik};
Dotted curve: Ambrosch-Draxl, Ref.~\cite{ambroschpr};
b.) Au, 1000 eV. Solid curve: Tung, Ref.~\cite{tungpr49};
Dash dotted curve: Palik, Ref.~\cite{palik};
Dotted curve: Hageman, Ref.~\cite{hageman};
}
\label{fcomp}
\end{figure}

These deviations, however, are of the same order of magnitude as differences between different sets of
optical data found in the literature, as shown in Figure~\ref{fcomp}. 
For Cu, the present data are in reasonable  agreement with the DIIMFP calculated from optical data in
Ref.~\cite{tungpr49} and Ref.~\cite{palik}, except for the two peaks at $\sim$20 and $\sim$30~eV which are slightly
sharper as expected from the optical data. The density-functional-calculations  of
Ref.~\cite{ambroschpr} agree closely with our data for the first peak at $\sim$20~eV, but deviate for higher
energies.  For Au, on the basis of the good mutual consistency of the present data,  and those of Ref.~\cite{tungpr49}
and Ref.~\cite{palik}, one might be inclined to suspect a systematic error in the dataset of Ref.~\cite{hageman}. This
comparison emphasizes the necessity of having an effective means for experimental determination of optical constants at
one's disposal.

Since, in principle, optical data can be derived from experimental DIIMFPs, the present
results  are encouraging in that the proposed approach seems to open up a new way to obtain such data. The
involved experimental procedure is conceivably straightforward and at any rate much simpler than other techniques to
obtain optical data in this energy range like light scattering or loss measurents in the transmission microscope. Most
importantly, data on the dielectric response can be derived from measurements with an inherent lateral resolution below a
few nm on a specimen  prepared in an arbitrary way (as long as it can be brought into an ultra-high vacuum
chamber). This constitutes an important advantage over other measurements including loss measurements conducted in
transmission.

A comparison as given here for Si, Cu and Au was made for other measured materials for which Drude--Lindhard
parameters for the dielectric function  have been published  (Ag, Al, Be, Bi, C, Co, Fe,
Ge, Mo, Ni, Pd, Pt, Ta, Te, Ti, V, W).   These comparisons showed similar features as those for
Si, Cu and Au, which therefore constitute a  representative subset of all results.

For materials with sharp  loss distributions  the energy/angular dependence of the DIIMFP/DSEP produced spurious
features in the retrieved data, as discussed for Si, and, in most cases, these were reproduced by the type III model
calculations in much detail. For the other materials these deviations were smaller. Nonetheless it seems worthwhile to
try to apply the proposed algorithm to angle resolved data (taken at the same energy) in order to reduce these
spurious features. The remarkable agreement between the type III model calculations and experimental retrievals
confirm the model of medium energy electron transport in solids in unprecedented detail (see Figure~\ref{flosssi}), in
particular the angular dependence of the {\em shape} of the  DSEP and the energy dependence of the {\em shape} of the DIIMFP, which
are very subtle effects.

An issue that deserves to be addressed at this stage is the influence of an uncertainty in the input parameters on the
outcome of the retrieval procedure. The input parameters comprise the elastic scattering cross section, and the total
scattering probabilities, i.e. the inelastic mean free path and the surface excitation probability, or average number
of surface excitations in a single surface crossing. While the scattering potential  employed to establish the elastic
cross  section is known to critically affect the angular distribution of the elastic peak intensity in some cases
\cite{powjabepes}, it is not expected to affect the sequence of {\em reduced} partial intensities in a significant way
since it is the {\em
shape}Êof the pathlength distribution that governs the latter, while the absolute elastic peak intensity  is determined
by the absolute value of pathlength distribution [cf. Equation~(\ref{epipld})]. The same
reasoning holds
for the inelastic mean free path that is used to convert the pathlength distribution into the sequence of partial
intensities.
These considerations are  supported by the fact that the bulk loss distributions  retrieved from experiment shown in
Figs.~\ref{flosssi}--\ref{flossau} agree quite well --on an absolute scale-- with the theoretical
normalized DIIMFPs.

The relative error in the surface partial intensities is proportional to the relative error in the surface
excitation parameter. the latter varies between $\sim$1--2 for all materials studied so far and is known with an
accuracy of about 10\% \cite{wersuxssl}.  Model calculations show that the main influence of an error in the SEP of
this order of magnitude on the outcome of a retrieval is that the area under the retrieved curve deviates from unity
by approximately the same relative amount. The deviations from theory observed for the DSEP of Cu can be explained in
this way.

\section{Summary} 
The energy loss process of electrons reflected from solid surfaces was studied theoretically. It was shown that the
Fourier transform of the loss spectrum can be written as a bivariate power series of the normalized  differential
surface and bulk loss distribution in individual collisions. A reversion of this bivariate power series is given on
the basis of two input spectra with sufficiently different sequences of partial intensities. The resulting procedure
was applied to model data and experimental spectra. The retrieved loss distributions compare satisfactorily
with theory. On the whole,  the results provide a detailed verification of the
commonly accepted model of medium energy electron transport in solids. In particular, the energy/angular dependence of
the {\em shape} of the DIIMFP and the DSEP was confirmed in detail for Si.

\section{Acknowledgments}
The author is grateful to Dr. Claudia Ambrosch-Draxl for making the optical data of Ref.~\cite{ambroschpr} available before publication and 
to   Drs. Helmut Werner and Werner Smekal for fruitful discussions. Financial support of the present work
by the Austrian Science Foundation FWF  through Project No. P15938-N02 is gratefully acknowledged


\newpage


\end{document}